\documentclass[12pt,preprint]{aastex}
\slugcomment{Submitted to ApJ Letters. May 9th 2006}

\shorttitle{Stellar spectra in M33}
\shortauthors{McConnachie et al.}

\begin{document}

\title{The stellar halo and outer disk of M33}

\author{Alan W. McConnachie$^1$\email{alan@uvic.ca}}
\author{Scott C. Chapman$^{1,2}$\email{schapman@submm.caltech.edu}}
\author{Rodrigo A. Ibata$^{3}$\email{ibata@astro.u-strasbg.fr}}
\author{Annette M. N. Ferguson$^{4}$\email{ferguson@roe.ac.uk}}
\author{Mike J. Irwin$^{5}$\email{mike@ast.cam.ac.uk}}
\author{Geraint F. Lewis$^{6}$\email{gfl@Physics.usyd.edu.au}}
\author{Nial R. Tanvir$^{7}$\email{nrt@star.herts.ac.uk}}

\affil{$^1$Department of Physics and Astronomy, University of Victoria, Victoria, B.C., V8P 1A1, Canada}
\affil{$^2$California Institute  of  Technology,   Pasadena,  CA  91125,  U.S.A.}
\affil{$^3$Observatoire   de  Strasbourg,   11,   rue  de l'Universite,   F-67000,  Strasbourg,  France}
  \affil{$^4$Institute for Astronomy, University of Edinburgh, Royal Observatory, Blackford Hill, Edinburgh EH9 3HJ, U.K.} 
\affil{$^5$Institute of Astronomy, Madingley  Road,  Cambridge, CB3  0HA,  U.K.}
\affil{$^6$Institute  of Astronomy, School  of Physics, A29, University of  Sydney, NSW 2006, Australia}
\affil{$^7$Physical Sciences,  University of Hertfordshire, Hatfield, AL10 9AB, U.K.\\}

\begin{abstract}
We present first results from a Keck/DEIMOS spectroscopic survey of
red giant branch (RGB) stars in M33. The radial velocity distributions
of the stars in our fields are well described by three Gaussian
components, corresponding to a candidate halo component with an
uncorrected radial velocity dispersion of $\sigma \simeq
50$\,km\,s$^{-1}$, a candidate disk component with a dispersion
$\sigma \simeq 16$\,km\,s$^{-1}$, and a third component offset from
the disk by $\sim 50$\,km\,s$^{-1}$, but for which the dispersion is
not well constrained. By comparing our data to a model of M33 based
upon its HI rotation curve, we find that the stellar disk is offset in
velocity by $\sim 25$\,km\,s$^{-1}$ from the HI disk, consistent with
the warping which exists between these components.  The spectroscopic
metallicity of the halo component is [Fe/H] $\simeq -1.5$,
significantly more metal-poor than the implied metallicity of the disk
population ([Fe/H] $\simeq -0.9$), which also has a broader colour
dispersion than the halo population. These data represent the first
detections of individual stars in the halo of M33 and, despite being
$\sim 10$ times less massive than M31 or the Milky Way, all three of
these disk galaxies have stellar halo components with a similar
metallicity. The color distribution of the third component is
different to the disk and the halo, but is similar to that expected for
a single, coeval, stellar population, and could represent a stellar
stream. More observations are required to determine the true nature of
this intriguing third kinematic component in M33.
\end{abstract}


\section{Introduction}

M33, the Triangulum Galaxy ($1^h 33^m 51^s, 30^\circ 39^\prime
36^{\prime\prime}$), is the third brightest Local Group galaxy, with
an integrated luminosity of $M_V = -18.9$. It is a late-type
spiral, Sc II-III, with two open spiral arms and no clear evidence of
any bulge component (\citealt{bothun1992,minniti1993,mclean1996}). The
overwhelming majority of light in M33 is distributed in an exponential
disk component (\citealt{devaucouleurs1959}), and the optical disk is
tilted at nearly $30^\circ$ to the strongly warped HI envelope
(\citealt{rogstad1976}). The mass-to-light ratio of its nucleus is
small ($M/L < 0.4$), ruling out the presence of a supermassive black
hole (\citealt{kormendy1993,lauer1998}) like that implied to exist in
the Milky Way (MW; \citealt{genzel1997}) and M31
(\citealt{dressler1988,kormendy1988}). 

\cite{chandar2002} analyse the kinematics of stars clusters in M33,
and find two distinct kinematic populations, which they argue
represent a disk and halo population, although individual stars
belonging to a halo component in M33 had not been directly
observed. In a companion paper by A. Ferguson et al. {\it in
preparation} (see also \citealt{ferguson2006}), we present our wide
field photometry of this galaxy taken with the Isaac Newton Telescope
Wide Field Camera (INT~WFC), and show that a low level stellar
component dominates the radial profile of M33 at large radius,
implying the presence of an extended stellar halo in M33. Unlike its
massive neighbor, M31, there is no evidence of substructure in M33 to
a surface brightness threshold of $\mu_V \simeq 30$\,mags\,sq.arcsec.

In this {\it Letter}, we present first results from a spectroscopic survey
of red giant branch (RGB) stars in M33 using the DEep Imaging
Multi-Object Spectrograph (DEIMOS) on Keck~II
(\citealt{faber2003}). Section~2 describes our observations. Section 3
analyzes the radial velocity distributions in our fields. Section
4 analyzes the metallicities of our target stars, and Section 5
discusses our results and concludes. We assume a distance to M33 of
$809 \pm 24$\,kpc (\citealt{mcconnachie2004a,mcconnachie2005a}).

\section{Observations}

\begin{figure}
  \begin{center}
    \includegraphics[angle=270, width=8cm]{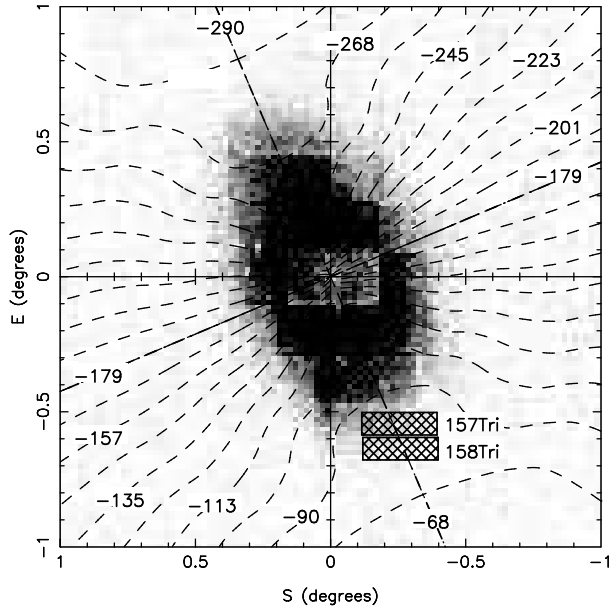}
    \caption{A tangent plane projection of RGB stars in M33 with $I <
    22$, from our INT WFC survey (Ferguson et al. 2006). The
    locations of the Keck/DEIMOS fields are indicated. Contours show
    the mean heliocentric radial velocity of a model M33 disk, based
    upon the Corbelli (2003) HI rotation curve.}
    \label{rotcurve}
  \end{center}  
\end{figure}

Since 2002, our group has been conducting a spectroscopic survey of
M31 using DEIMOS on Keck~II
(\citealt{ibata2004,ibata2005,mcconnachie2004b,chapman2005,chapman2006}). In
September -- October 2005, we extended the scope of this survey to
M33. The two fields analyzed here were taken at the end of the night
at high airmass, during imperfect weather conditions. Their location
relative to the disk of M33 can be seen in Figure~\ref{rotcurve}, and
are positioned $\sim 38^\prime$ (9\,kpc) along the southern major axis
to study the kinematics of stars at the edge of the M33 disk.  In the
same way as our M31 study, we used the `mini-slitlet' DEIMOS setup for
these fields, which significantly increases the multipexing
capabilites of this instrument
(\citealt{ibata2005,chapman2006}). Targets were selected from the INT
survey using a color -- magnitude box defined by $20.5 < i < 22.0$ and
$1.0 < \left(V - i\right)_o < 4.0$. Other targets brighter than $i =
22$ were automatically chosen to fill in the available space of the
spectrograph detector.

The contours in Figure~\ref{rotcurve} show the expected mean stellar
heliocentric radial velocities for a kinematic model of M33, based
upon the HI rotation curve of M33 derived by \cite{corbelli2003} and
shown in her Figure~5. We assume a stellar distribution in M33
described by $\rho\left(R, z\right) \propto
\exp\left(-R/R_e\right)\exp\left(-z/z_e\right)$, with $R_e =
5.8^\prime$ (1.4\,kpc; \citealt{regan1994}) and $z_e =
0.2$\,kpc. Changing $z_e$ within the range $\left[0,1\right]$\,kpc
makes no significant difference to any of our results.  Stars occupy
circular orbits with $v_z = 0$ and $v_R$ defined by interpolation of
the HI rotation curve, linearly extrapolated beyond $R = 77^\prime$
(18\,kpc) where necessary. The model is then corrected for the
position angle ($\theta \simeq 23^\circ$) and inclination ($i \simeq
54^\circ$) of the stellar disk of M33. The mean velocity along any
sight-line through the disk is calculated as $\bar{v_l} = \int
\mathbf{v} \bullet \hat{\mathbf{L}}~\rho(R, z)~dl / \int \rho(R,
z)~dl$, and corrected to the heliocentric frame. $\hat{\mathbf {L}}$
is a unit vector pointing along the direction of our line of sight.

The data were processed and reduced in the same way as the main M31
survey (\citealt{ibata2005,chapman2006}). Due to the generally poor
observing conditions, the data analyzed here consist of $280$\,stars
with a cross correlation coefficient $> 0.02$, a signal-to-noise
($S/N$) in the continuum $> 2$, a Tonry-Davis coefficient
(\citealt{tonry1979}) $> 2$, and a velocity error $v_{err} <
25$\,km\,s$^{-1}$ ($\overline{v}_{err} \simeq 10$\,km\,s$^{-1}$).

\section{Kinematics}

\begin{figure}
  \begin{center}
    \includegraphics[angle=0, width=8cm]{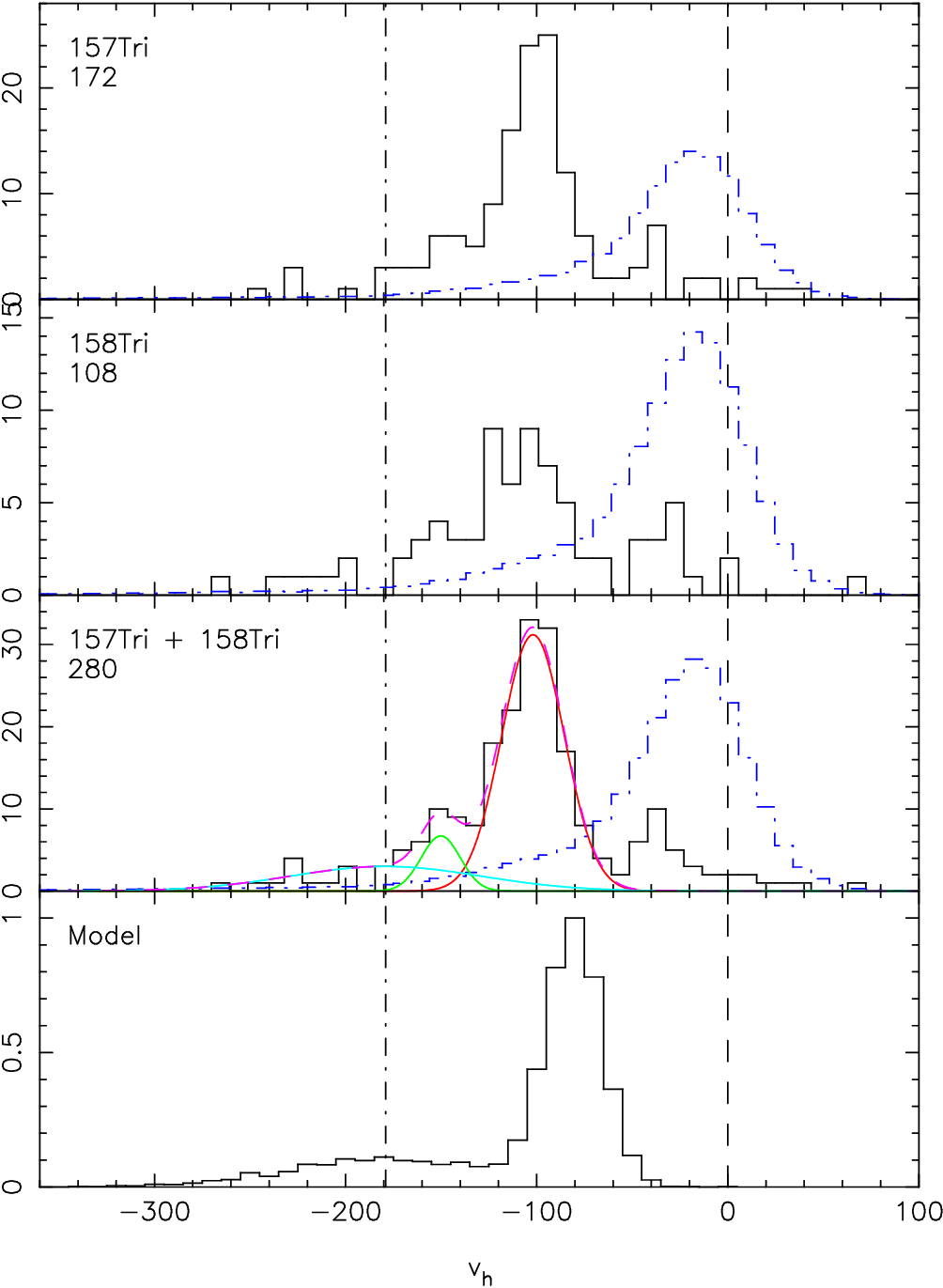}
    \caption{First and second panels: heliocentric radial velocity
    distributions for our M33 fields. The dashed lines represent zero
    velocity, and the dot-dashed line represents the systemic velocity
    of M33, $v_h = -179$\,km\,s$^{-1}$. The dot-dashed histograms show
    the heliocentric velocity distribution of the expected MW
    foreground stars in each field, from the MW model by
    \citealt{robin2003}, which satisfy our color selection
    criteria. These histograms have been scaled to match the area of
    the DEIMOS fields, and the small number of stars at $v_h >
    -80$\,km\,s$^{-1}$, which are most likely foreground, suggest that
    contamination at more negative velocity will be negligible. Third
    panel: Same as first two panels, except now the two fields have
    been added together to boost statistics. Three Gaussian velocity
    distributions and their summed total are overlaid. See text for
    details. Fourth panel: the expected heliocentric radial velocity
    distribution of stars in our fields, using our M33 disk model and
    an additional Gaussian component, representing a smooth stellar
    halo.}
    \label{3fields}
  \end{center}
\end{figure}

The top two panels of Figure~\ref{3fields} shows the heliocentric
radial velocity ($v_h$) distributions in our two fields. The dashed
lines indicate $v_h = 0$, and the dot-dashed lines indicate the
systemic velocity of M33, $v_h = -179$\,km\,s$^{-1}$. The dot-dashed
histograms show the expected heliocentric velocity distributions of
the MW stars that lie in each field using the MW model of
\cite{robin2003}, and which satisfy our selection criteria for M33 RGB
stars. The histograms have been scaled to match the area of the DEIMOS
fields. They show that any contribution from MW foreground stars at
$v_h < -80$\,km\,s$^{-1}$ is small for the purposes of this {\it
Letter}, and so this velocity cut is adopted in the following
kinematic analysis. Stars with $v_h > -80$\,km\,s$^{-1}$ are
predominantly MW foreground, and the small numbers we observe further
suggest that the contamination at more negative velocity will be
negligible.

The third panel of Figure~\ref{3fields} shows the velocity
distributions of Fields 157 and 158 added together to boost
statistics. We expect the main stellar components in M33 will be well
fit by individual Gaussians, and any stellar halo component in M33
will have a velocity distribution centered on $v_h =
-179$\,km\,s$^{-1}$. Since M33 is $\sim 10$\,times less massive than
M31, we expect that the M33 halo velocity dispersion will be $\sim
\sqrt{10}$ that of M31 ($\sigma_{M31} \sim 145$\,km\,s$^{-1}$;
\citealt{chapman2006}). The broad Gaussian overlaid in the third panel
of Figure~\ref{3fields} represents a M33 halo component with a
dispersion of $\sigma = 50$\,km\,s$^{-1}$, scaled to match the
approximate number of stars observed. We also expect that the
prominent peak of stars in these fields is due to the dominant disk of
M33. Therefore, we have subtracted the halo Gaussian from the $v_h$
distribution in the third panel and fitted a Gaussian to this peak,
centered at $v_1 = -102 \pm 2$\,km\,s$^{-1}$ with an uncorrected
dispersion $\sigma_1 = 16 \pm 3$\,km\,s$^{-1}$ (corrected dispersion
of $\sim 12.5$\,km\,s$^{-1}$). The residuals of this fit show a
prominent grouping of stars at $v \sim -150$\,km\,s$^{-1}$. We fit a
third Gaussian to this feature, with a peak $v_2 = -150 \pm
4$\,km\,s$^{-1}$ and a dispersion $\sigma_2 = 10 \pm 4$\,km\,s$^{-1}$.
The latter quantity is more poorly constrained than given by the
formal error, however, as its dispersion is covariant with the
dispersion of the disk and halo, which have been fitted
independently. All 3 Gaussians and their summed total are overlaid in
the third panel of Figure~\ref{3fields}.

The fourth panel of Figure~\ref{3fields} shows the expected
distribution of stars at the positions of Fields 157 and 158, for the
model of M33 shown in Figure~\ref{rotcurve}, combined with a halo
model with a radial velocity dispersion of $50$\,km\,s$^{-1}$. The
radial velocity dispersion of the disk is set to $16$\,km\,s$^{-1}$,
to mimic our observations. The normalization of both components is
chosen to approximately match the corresponding peaks in the data . A
comparison between model and data shows (1) the peak at $v_2 =
-150$\,km\,s$^{-1}$ does not naturally arise as a result of our
line-of-sight through the main M33 disk sampling stars at different
circular velocities, (2) the stellar disk is offset in $v_h$ from the
HI disk by $\sim 20 - 25$\,km\,s$^{-1}$, implying a deprojected offset
of $\sim 25 - 30$\,km\,s$^{-1}$. For comparison, the asymmetric drift
between the stars and HI in the MW and M31 is of order $15 -
20$\,km\,s$^{-1}$. Scaling for mass, the expected asymmetric drift in
M33 is not likely to be more than $\sim 10$\,km\,s$^{-1}$. We believe
that the velocity offset between the two disks is due to differential
warping, specifically that the HI disk is warping into our
line-of-sight at large radius from M33. This has been implied, quite
spectacularly, by \cite{rogstad1976} (see their Figure~14), and is
consistent with the magnitude of the velocity offset.

\section{Metallicities}

\begin{figure*}
  \begin{center}
    \includegraphics[angle=270, width=15cm]{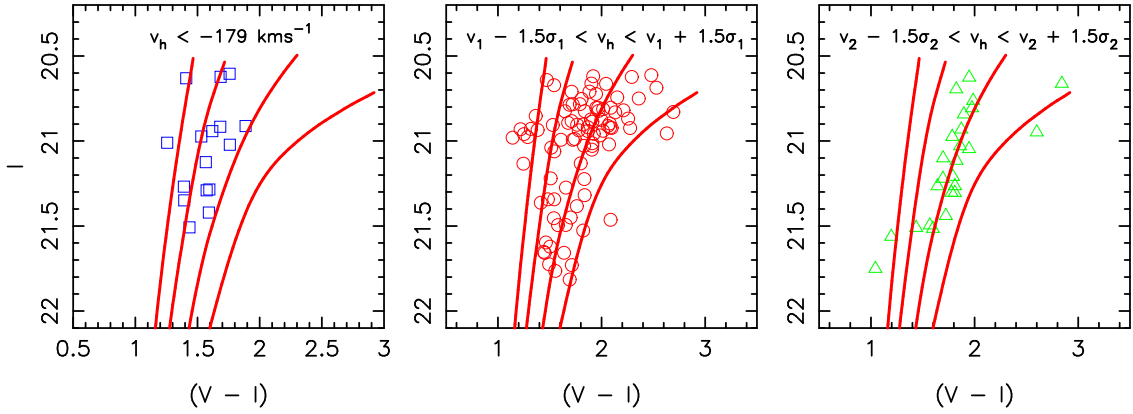}
    \caption{CMDs of spectroscopically observed RGB stars from Fields
    157 and 158, selected to satisfy the velocity cuts indicated
    ($\sigma_1 = 16$\,km\,s$^{-1}$, $\sigma_2 =
    10$\,km\,s$^{-1}$). These velocity cuts predominantly sample stars
    which belong to the candidate halo (left panel), disk (middle
    panel) and the `third component' (right panel). Victoria -- Regina
    isochrones (VandenBerg et al. 2005), corresponding to 13\,Gyr old
    stars and with metallicities of [Fe/H] $= -2.0, -1.3, -0.8$ and
    $-0.5$, are overlaid.}
    \label{cmd}
  \end{center}
\end{figure*}

Figure~\ref{cmd} shows the color -- magnitude diagrams (CMDs) for the
spectroscopically observed M33 RGB stars in Fields 157 and 158,
divided by the radial velocity cuts indicated so as to kinematically
separate the three candidate components as best as possible. Victoria
-- Regina isochrones from \cite{vandenberg2005}, with $BVRI$ color --
$T_{eff}$ relations as described by \cite{vandenberg2003},
corresponding to a 13 Gyr stellar population with metallicities of
[Fe/H] $= -2.0, -1.3, -0.8$ and $-0.5$, are also overlaid.

The differences between the CMDs in Figure~\ref{cmd} are striking, and
appear to validate our kinematic deconvolution. The candidate M33 halo
stars are predominantly bluer and have a smaller color dispersion than
the candidate disk stars, indicating a difference in the age and/or
metallicity properties of the populations. Assuming both populations
are $13$\,Gyrs old implies a median photometric metallicity for the
halo of [Fe/H] $= -1.3 \pm 0.1$, with a dispersion of $\sigma_{{\rm
[Fe/H]}} = 0.21$. Given the small number of stars which contribute to
this measurement, this is consistent with the Ferguson et al. (2006)
metallicity measurement for the halo, of [Fe/H] $= -1.5 \pm 0.1$. The
disk has an implied photometric metallicity of [Fe/H] $= -0.9 \pm 0.1$
and a broader dispersion of $\sigma_{{\rm [Fe/H]}} = 0.35$. The median
color (metallicity) of the `third component' is similar to the disk,
although its color dispersion is remarkably narrow ($\sigma_{{\rm
[Fe/H]}} = 0.15$). This population is difficult to isolate
kinematically, and our sample could be contaminated by both of the
other two components. However, the locus of its stars on the CMD does
not clearly associate it with either the candidate disk or halo
population, and is reminiscent of a CMD for a single stellar
population. If this population was due to MW foreground contamination,
we would not expect its stars to occupy so well defined a locus in
this diagram. The most probable explanation is that these stars belong
to a stellar stream. These stars do not, however, define a spatially
obvious structure, but given our limited field size and that our
choice of slitlet position will affect the spatial distribution, this
test is inconclusive at best.

We calculate the spectroscopic metallicity of the halo stars by
measuring the equivalent width of the CaT from stacked spectra, using
an identical method to \cite{ibata2005} and \cite{chapman2006}. We
find that [Fe/H] $= -1.5 \pm 0.3$, in good agreement with the
photometric measurement. The large uncertainty reflects the
relatively low signal-to-noise of the stacked spectrum. We are unable
to measure the metallicities of the other two components in this way,
since their velocities cause the CaT to lie on top of strong sky lines
which prevents accurate equivalent width measurements.

\section{Discussion}
\label{final}

M33 is considered an example of the quintessential disk galaxy, with
no evidence for any bulge or spheroid component. However, the radial
profile for M33 presented in Ferguson et al. (2006), based upon our
INT-WFC observations, reveals for the first time the presence of a
spatially extended, low density component. Using Keck/DEIMOS, we have
identified a population of RGB stars which have velocities centered on
the systemic velocity of M33 and which appear to have a large
dispersion ($\sigma \sim 50$\,km\,s$^{-1}$). These stars are bluer
than other RGB stars in M33, with a small color dispersion. Their
average metallicity is [Fe/H] $\sim -1.5$, and we attribute them to
the newly discovered halo component of M33. The haloes of the MW and
M31 have similar metallicities to this
(\citealt{eggen1962,chapman2006,kalirai2006b}), even although both
these galaxies are $\sim 10$\,times more massive than M33. It is
important to emphasise that there is no evidence of a trend linking
halo metallicity to galaxy mass or luminosity for the Local group
galaxies, as has been suggested from recent Hubble Space Telescope
studies of galaxies within $\sim 10$\,Mpc
(\citealt{mouhcine2005a,mouhcine2005b}). Instead, all the large Local
Group spiral galaxies appear to have very similar halo metallicities.

We also observe stars in the disk of M33. Their $v_h$ distribution is
well described by a Gaussian with an uncorrected dispersion of $\sigma
\simeq 16$\,km\,s$^{-1}$, corresponding to a corrected dispersion of
$\simeq 12.5$\,km\,s$^{-1}$. These stars are redder and have a broader
color dispersion than the halo stars. The stellar disk is offset in
$v_h$ from the HI disk in our simple model by $\sim 20 -
25$\,km\,s$^{-1}$, which is fully consistent with the extreme warping
of the HI disk implied by \cite{rogstad1976}.

Finally, we present evidence for a third kinematic feature in M33,
revealed by the $v_h$ distribution of the southern fields. The
velocity dispersion of this component is not well constrained by our
data. A comparison to our simple model of the disk of M33 shows that
this feature does not arise as a result of our line-of-sight through
M33 sampling stars at different circular velocities in the main
disk. The locus of these stars in the CMD has a remarkably small color
dispersion, reminiscent of a single stellar population and
inconsistent with belonging to a foreground population. We believe the
most likely explanation for this feature is that it is a stellar
stream. However, more spectroscopic observations of fields at
different locations around the disk of M33 are required before the
true nature of this intriguing feature is revealed.

\acknowledgments The data presented herein were obtained at the
W.M. Keck Observatory, which is operated as a scientific partnership
among the California Institute of Technology, the University of
California and the National Aeronautics and Space Administration. The
Observatory was made possible by the generous financial support of the
W.M. Keck Foundation. We thank E.~Corbelli for supplying us with her
data for the HI rotation curve of M33, and the anonymous referee for
constructuve comments . AWM would like to thank J.~Navarro and
S.~Ellison for financial support. AMNF is supported by a Marie Curie
Excellence Grant from the European Commission under contract
MCEXT-CT-2005-025869. GFL acknowledges support through ARC DP0343508.

\clearpage

\clearpage

\clearpage

\end{document}